# Degeneracy breaking of Wood's anomaly for enhanced refractive index sensing

Michal Eitan[†,‡,∥], Zeev Iluz[§,∥], Yuval Yifat[†,‡,∥], Amir Boag[†], Yael Hanein[†,‡], and Jacob Scheuer[†,‡,*]

[†] Department of Physical Electronics, School of Electrical Engineering, Tel-Aviv University, Tel-Aviv 69978, Israel

[‡] Tel-Aviv University Center for Nanoscience and Nanotechnology, Tel-Aviv University, Tel Aviv 69978, Israel

[§] CST AG, Bad Nauheimer st. 19, 64289 Darmstadt, Germany

ABSTRACT: We introduce an ultrasensitive detection technique for refractive index (RI) sensing based on an array of nanometer scale slot-antennas milled in a thin gold layer using a single lithographic step. Our experimental figures of merit (FOMs) of 140-210 in the telecom wavelength range approach the fundamental limit for standard propagating SPR sensors (~250). The underlying mechanism enabling this is the combination of a narrowband resonance of the slot-antennas with degeneracy breaking of Wood's anomaly under slightly non-perpendicular illumination. In addition, we explore the sensitivity of the device to the depth of the analyte layer. This concept can be easily tuned to any desired wavelength and RI range by modifying the slot dimensions and the array spacing, thus rendering it highly useful for numerous sensing applications.

In the last decade Surface plasmon resonance (SPR) effects have received great attention due to their attractiveness for a broad range of applications such as surface-enhanced Raman scattering[1], wide-angle holography[2], bio-molecular sensing[3–6], solar power harvesting[7], and sub-wavelength optical imaging[8]. Among these applications, plasmonic based, label-free, nano-sensors (in terms of interaction volume) have been at the focus of numerous studies, utilizing diverse structures such as nano-particles, nano-antennas, and meta-surfaces, to name just a few examples[9–20]. SPR sensors usually exploit the extremely high sensitivity of the spectral response of plasmonic structures to their close surroundings in order to detect minuscule changes in the ambient RI. Such sensors can be roughly classified as either localized or propagating SPR devices, based on whether the excited plasmon is localized to a metallic nano sized structure, or propagates along an interface. Conventional propagating SPR sensors can be excited in a variety of illumination schemes but the most simple and efficient design, the Kretschmann configuration, is based on a standard prism coupled to a metallic surface. While localized SPR typically exhibit substantially lower sensitivities (up to two orders of magnitude) as compared to the propagating SPR devices, their simplicity, low-cost, compact dimensions, and suitability for integration, make them highly attractive for practical devices, a fact which has triggered substantial efforts aimed towards enhancing their performance.

The generally accepted metrics for evaluating the efficiency of RI sensors are their sensitivity, defined as a ratio of the resonance wavelength shift to the change in the refractive index, namely $S = \Delta\lambda_p / \Delta n_{ambient}$, and its FOM which measures the sensitivity relative to the full width at half maximum (FWHM) of the measured peak: $FOM = S/FWHM$. As the FOM metric provides a gauge for the minimal detectable change in the RI, it is, in many cases, more important than the sensitivity, S.

Past studies utilizing localized SPR employed one or both of the following mechanisms to enhance the sensitivity and FOM: 1) Coupling bright and dark SPRs by means of employing two plasmonic nano-structures or geometrical symmetry breaking in order to obtain a Fano-shape[21–24] or an EIT (electromagnetically induced transparency)-like narrow linewidth resonance via the mechanism of plasmon hybridization[25]; 2) Increasing as much as possible the interface between the metallic nano-structure and the surrounding medium by, e.g. placing them on a dielectric nano-pillar to reduce the effect of the substrate on the fields generated at the plasmon resonance[26]. Despite these efforts, the FOM values of localized SPR remain lower than those of propagating SPR. Owing to their attractiveness as compact sensors, there is a clear interest in further improving the performances of localized SPR sensors.

In this letter, we present and experimentally demonstrate a new proof of concept for achieving high FOM using narrow slot antenna array (NSAA) milled in a thick layer of gold. In addition, we numerically analyze the sensitivity dependence of the device on the thickness of the analyte layer and find that it maintains its high FOM values even for layers below a single wavelength.

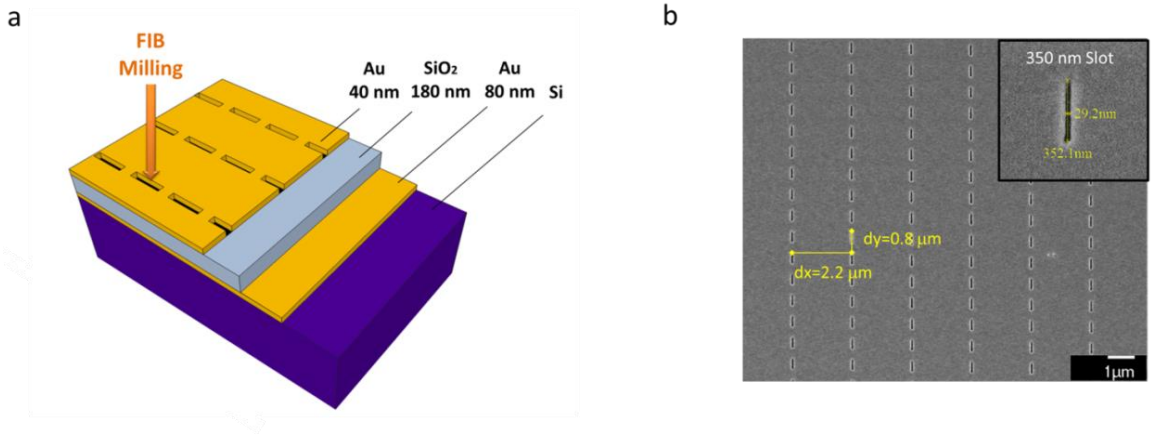

Fig. 1. NSAA structure and fabrication. (a) Schematic of device structure. (b) SEM image of an array of 350 nm long slot nano-antennas with unit cell size of dx=2.2 µm and dy=0.8 µm. Inset: a zoom-in image of an individual slot antenna.

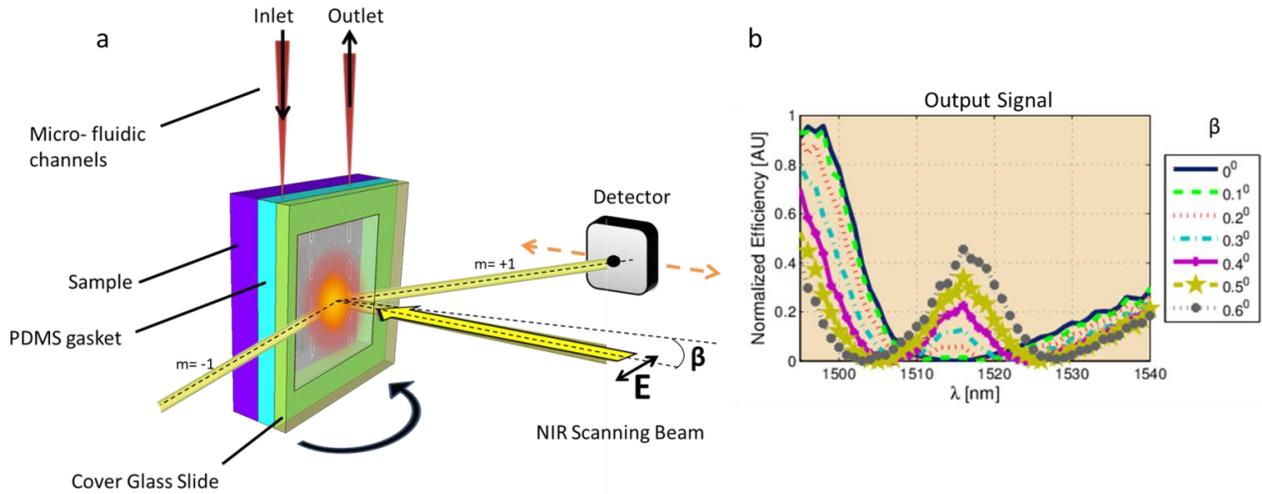

Figure 2. NSAA characterization setup (a) The NSAA is placed in a sealed micro-chamber (a glass cover and a PDMS gasket) into which the analyte is inserted. Linearly polarized light from a tunable laser source is then directed onto it and the spectral properties of the m=+1 Bragg lobe is measured. At each tilt angle, β, the detector was repositioned in order to track the scattering angle of the grating lobe. (b) The output - measured spectral response of the 1st scattered lobe for different tilting angles (0°-0.6°) when the NSAA is immersed in isopropanol (IPA).

Figure 1 shows the NSAA structure used in this investigation. The length and width of each slot are approximately 350 nm and 30 nm, respectively. The slot antennas are milled in a 40 nm thick gold layer deposited above a 180 nm SiO2 buffer layer and an 80 nm Au backplane reflector. Such antennas are equivalent to dipole antennas over a dielectric substrate through Babinet's principle[27], and are characterized by a narrowband spectral response which can be tuned over a large range of operating wavelengths by modifying the slot dimensions and the array spacing. Compared to the more commonly employed geometry of dipole antenna array, the NSAA resonance is substantially sharper. In addition, fabricating ultra-narrow slot antennas (using, e.g. focused ion beam milling) yields, generally, higher resolution lithography compared to the equivalent dipole antennas. An SEM image of the structure is depicted in Fig. 1b. The array spacing is chosen such that it generates only the first order Bragg diffraction lobe in air over the measured wavelength range (1460-1640 nm). This diffraction lobe serves as the optical output signal of the device, allowing for highly sensitive measurement of the NSAA spectral response and consequently, any changes in the ambient RI[32]. The high sensitivity stems from the properties of the modal field profile of slot-antenna, which is concentrated in the vacancy of the structure and, therefore, is strongly affected by the RI of the surrounding medium.

The optical properties of the NSAA were characterized using a far field setup described schematically in Fig. 2a and is based on that used in Ref. [32]. Linearly polarized light is spectrally swept and collimated towards the NSAA, which is placed in a microfluidic chamber filled with a liquid. A detector is placed in the far field to measure the power scattered to the m=+1 lobe.

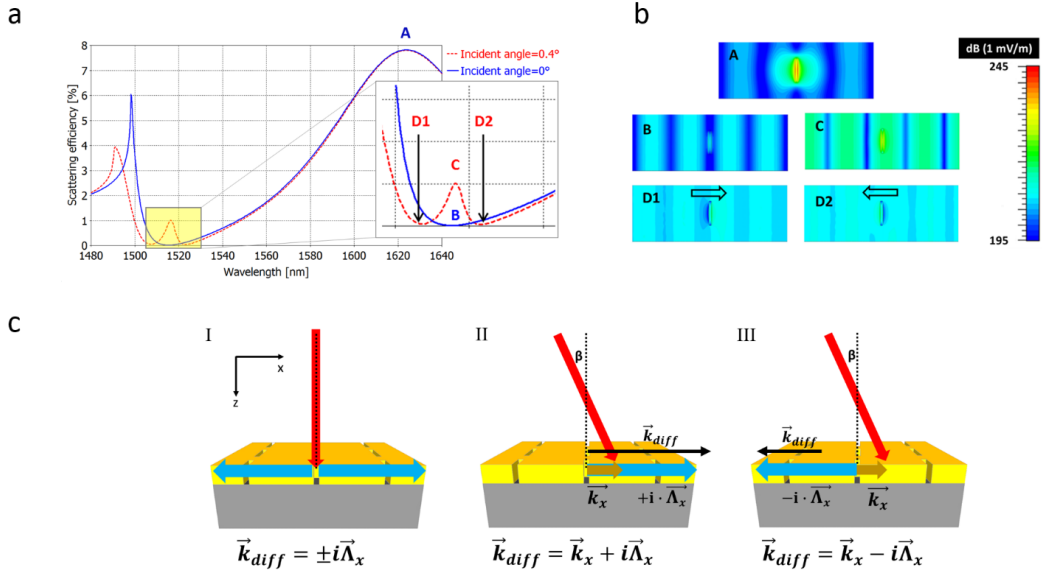

Figure 3. Formation of a sharp peak resulting from geometrical symmetry breaking when the incident beam is tilted by a small angle. (a) Simulated scattering efficiency of the first-order Bragg lobe for incident angle of 0° and 0.4° at RI of $n=1.362$. (b) Snapshots of the simulated propagating E-fields on the gold surface for the points marked in panel (a) (see also the temporal evolution in the supporting information). (c) WA formation and wavelength splitting: (I) Phase matching condition when the incident light is normal to the surface. When the light is incident at an angle of β, a right-propagating (II) and a left-propagating (III) waves are appearing at different wavelengths.

A representative example of the spectral properties of the power scattered to the m=+1 Bragg lobe for various incidence angles (close to normal) is depicted in Fig. 2b. An important feature shown in the figure is the formation of almost a null in the scattered power at 1512 nm for normal incident which is replaced by a narrow peak when angle of the incident field is slightly tilted. The formation of this peak is a manifestation of the physical phenomenon that our sensing scheme is based on.

The underlying physical mechanism utilized by the device is the Wood's anomaly (WA) - a resonant reduction in the scattered power from a periodic structure caused by the diffraction of the incident light into a right-propagating and left-propagating surface waves parallel to a grating surface[11,28]. When the NSAA is tilted at a small angle β with respect to the wave-vector of the incident light, two distinct notches appear in the spectrum of the m=+1 Bragg lobe (see Fig. 2b). These notches correspond to the left and right propagating surface waves of the WA which at this configuration are excited at different wavelengths. The scattering peak which is formed between these notches due to the wavelength degeneracy removal of the left and right propagating surface waves (at tilted illumination) can be quite narrow (~5 nm FWHM in our device). As detailed below, the combination of the high sensitivity of the WA to the ambient refractive index (~1060 nm/RIU) and the narrow linewidth induced by the symmetry breaking lead to high FOM values (140-210) which approach the theoretical limit predicted for propagating SPR sensing in the telecom wavelength[29].

To further understand the emergence of the peak at tilted illumination angles as well as its utilization for sensitive RI measurement, we revisit the WA phenomenon in metallic grating. In such grating structure, comprising a metallic layer with a (complex) dielectric coefficient $\varepsilon_1$ surrounded by an analyte with a dielectric coefficient of $\varepsilon_2$, the surface waves corresponding to WA are excited when the momenta of the incident and surface waves are matched by the grating[28,33,34]

$$\vec{k}_{diff} = \vec{k}_x + l\vec{\Lambda}_x + m\vec{\Lambda}_y \quad (1),$$

where $\vec{k}_{diff}$ is the wavevector of the excited surface wave, $|\vec{k}_{diff}| = \omega/c\sqrt{\varepsilon_1\varepsilon_2/(\varepsilon_1+\varepsilon_2)}$, $\vec{\Lambda}_x = 2\pi/d_x\hat{x}$ and $\vec{\Lambda}_y = 2\pi/d_y\hat{y}$ are the reciprocal lattice vectors of the grating in the x and y directions, respectively. Here, $\vec{k}_x = \hat{x}k_0\sin\beta$ is the wavevector component of the incident light which is parallel to the surface (see Fig. 3c), and $l, m = 0, \pm 1, \pm 2, \ldots$. Additionally, we emphasize that in our treatment, β is the angle of the impinging light with respect to the normal to the surface in air. Note that when the array is surrounded by a different RI with n≠1 this is not the physical angle at which the incident beam impinges upon the metallic grating (the actual incident angle in this case is determined by Snell's law). $\lambda_0$ is the wavelength in vacuum and the integers l and m indicate the diffraction order along the x and y directions of the array, respectively. In our structure, the grating spacing in the y direction is substantially smaller than the wavelengths of interest, thus preventing the generation of Bragg lobes in this direction. Consequently, we set m=0. When the incident light impinges normally to the surface, i.e. β=0, the wavelength at which the WA occurs satisfies:

$$\vec{k}_{diff} = l\vec{\Lambda}_x \quad (2).$$

The solution of Eq. (2) yields the well-known equation for coupling to a surface plasmon via a diffraction grating[33,35] $l\lambda_0 = d_x\sqrt{\varepsilon_1\varepsilon_2/(\varepsilon_1+\varepsilon_2)}$. The positive and negative vectorial solutions correspond to the wavelength-degenerate rightwards and leftwards propagating surfaces waves which are simultaneously and equally excited. However, when the light is incident at a small, non-zero angle, this

degeneracy is removed and the grating equation exhibits two solutions for the WA:

$$\vec{k}_{diff} = \hat{x}k_0 \sin\beta \pm l\vec{\Lambda}_x \qquad (3),$$

The two distinct solutions of Eq. (3) correspond to the excitation of surface waves in each direction, which occur at different wavelengths because of the symmetry breaking. The left-propagating and right-propagating surface waves are excited at:

$$l\lambda_0^{\pm} = d_x\left(\sqrt{\varepsilon_1\varepsilon_2/(\varepsilon_1+\varepsilon_2)} \pm \sin\beta\right). \qquad (4),$$

The momentum conservation conditions at the normal and tilted incident beam configurations are depicted in Fig. 3c.

The splitting phenomenon was also explored by numerically simulating the scattering process from the NSAA structure using finite element frequency domain solver, (CST MWS[36]). Figure 3a shows the simulated spectral response of the 1st order scattered lobe from the NSAA structure when it is immersed in IPA ($n$=1.362), and illuminated at β=0° and 0.4° (in air). The simulated fields on the surface of the NSAA at selected wavelengths (defined in Fig. 3) are shown in Figure 3b.

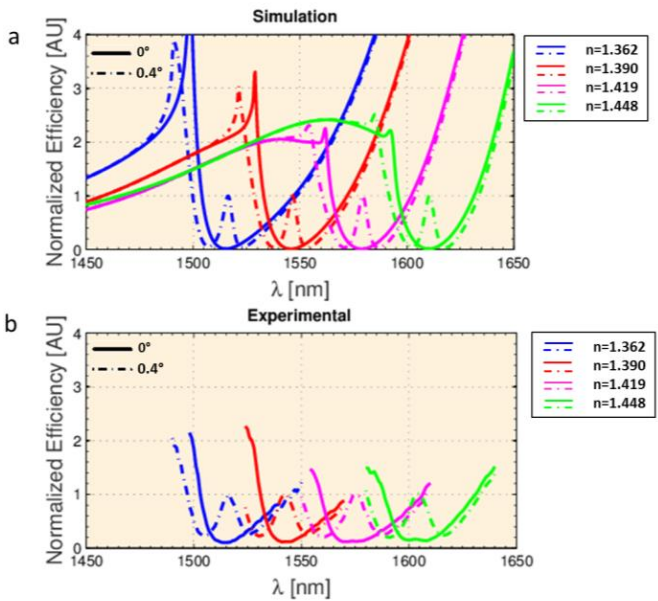

Figure 4. (a) Simulated and (b) measured scattering efficiencies versus wavelength for a sample illuminated at β=0° and β=0.4° relative to the incident beam. The results are shown for mixed blends of IPA and Cargille AA brand refractive index matching fluid with RI values of n=1.362 (blue), 1.390 (red), 1.419 (pink), and 1.448 (green).

To test the sensing performance of these devices, the fabricated devices were placed in a fluidic chamber as shown in Fig. 2a. The NSAAs were illuminated using collimated beam emitted by a tunable laser source in the wavelength range of 1440–1640 nm. Spectral responses of NSAAs were measured in mixed blends of IPA and Cargille AA brand refractive index matching fluid with RI values of n=1.362, 1.390, 1.405, 1.42, and 1.448 and at incidence angles between 0° (normal) and 0.6°.

Fig. 4 shows the experimental and numerically calculated scattering efficiencies (defined as the percentage of light scattered to the measured grating lobe relative to the total incident power) for incidence angles of 0° and 0.4°. A very good agreement is found between the theoretical and experimental results in terms of the position and width of the notches associated with WA and the sharp peaks for all RI values. The experimentally measured and the simulated scattering efficiencies have been individually normalized to counteract the effect of loss stemming from wavelength dependent absorption in the medium. The slight difference between the measured locations of the nulls and the calculated values is attributed to an inaccurate estimation of the RI values of the IPA - index matching fluids mixtures, possibly due to trace amounts of residual material left in the chamber during the experiment and slight deviations in the ambient temperature of the analyte from 25°C (the nominal temperature at which the RI of the materials used in the experiments were obtained). In addition, the absorption loss of the analyte was not taken into account in the simulations. The combination of these factors induce slight deviations in analyte RI[37].

The deviation in the RI value also explains the minor difference between the experimentally obtained and the theoretically calculated FOM, as discussed below. We also note that the normalized reflectance at the measured "null" did not reach zero as it does in the corresponding calculated results. This is likely due to the grainy nature of the Au surface which hinder the propagation of the surface waves, and slight imperfections in the shape and location of the individual antennas. These factors result in a decrease of the coupling to a surface wave and result in diffusive reflection. The complete set of measurements for all 5 RI liquids for angles 0° - 0.6° is shown in Fig. S2 of the supporting information.

Fig. 5 demonstrates the experimental sensitivity and FOM values obtained by analyzing the spectral results for all analytes, at incidence angles of 0.2° to 0.6°. For a tilt angle of 0.2°, the measured sensitivity and FOM were found to be as high as 1060 nm/RIU and 208, respectively. The sensitivity levels of the device were found to be independent of the tilt angle, however, the measured FOM values increased from approximately 100 to more than 200 as the incident tilt angle decreased. The sensitivity and FOM results for different tilt angles along with a detailed explanation of the method used for calculating the FWHM of the system are presented in the supporting information. Concurrently, numerical simulations show that the maximal FOM value for the studied device is 225, attained at a tilt angle of 0.25°, thus, approaching the fundamental limit for standard propagating SPR sensors (~250) which was calculated from Svedendahl et al.[29] (see also in the supporting information).

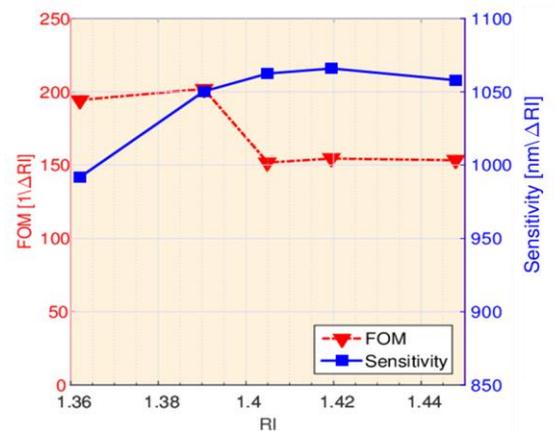

Figure 5. FOM and sensitivity values for the NSAA sensor versus different ambient RIs for mixed blends at illumination angle of 0.2°.

The measured sensitivity, *S*, and FOM allow for the evaluation of the minimal detectable RI limit (DL) given by[38]:

$$\text{DL} = \frac{R}{S} \quad (5),$$

where *R* is the sensor resolution defined as the smallest possible spectral shift that can be accurately measured. In the measurements described above, the spectral response of the slot antenna array was obtained by scanning the wavelength of the incident beam in 1 nm steps. This value was arbitrarily chosen to simplify the measurement and to reduce measurement times. However, as the spectral width of the tunable laser which was used is significantly smaller (on the order of ~1 pm), the main limitation for the sensor resolution does not stem from the laser linewidth, nor from the scanning step size, but rather from the ability to accurately extract the location of the central peak from the measured data. In our current configuration this ability is limited by the noise of the detector which at the lowest measured value is as high as 5% of the measured signal, resulting in an upper bound of DL~$5\times10^{-5}$. This limit can be easily improved by increasing the incident optical power (above 30 μW), thereby improving the signal to noise ratio (SNR) at the detector. As a result, minimal detectable RI values in the range of DL~$1\times10^{-6}$ can readily be achieved. Further improvement can be obtained by using different configuration, in which the light is directed through the substrate, rather than through the analyte.

The suitability of the NSAA sensor for bio-medical and small molecules sensing was explored by numerically simulating the optical response of the sensor tilted at 0.4 degrees to analyte layers of varying thickness with a fixed overall medium thickness. Figure 6 presents the sensitivity results for an n=1.45 layer of various thicknesses covered by a thick layer of n=1.4. The plots were obtained by tracking the spectral shifts of the sharp peak induced within the WA notch, and that of the slot antenna resonance. See supporting information for plots of the NSAA spectral properties for different analyte layer thickness. The sensitivity of the WA based method approaches the bulk level for layer thickness around 1000 nm (which is less than a single wavelength in the medium). This result is higher than previously reported results which were obtained for different wavelength ranges[39–41]. Note, that by tracking the shift of the slot antenna resonance, one can obtain reasonably high sensitivity values even for analyte layers of 100 nm although the corresponding FOM is not as high.

The reason for the different sensitivities of the antenna peak and the WA mechanisms is due to the difference in field distribution for the two phenomena. When the antenna is excited at resonance, a LPSR is excited around the antenna, resulting in a spatially concentrated field which is sensitive to the immediate environment around the antenna. As a result, the antenna resonance yields a much higher sensitivity to low thickness analyte layers, approaching peak sensitivity for layers of 100 nm. Conversely, when a WA is excited, the result is a propagating surface wave, which has a longer range than the LPSR. The WA effect which has a more distributed field, will sense an area on the order of the travelling wavelength, and as a result, will be less sensitive to sub wavelength changes in the analyte. Simulated field maps of both cases can be seen in the supporting information. Nonetheless, both sensing mechanisms show a promise for biomedical applications which require sensitivities to changes in the order of 1 μm [42,43] and below[30], and we plan to experimentally verify this in the future research.

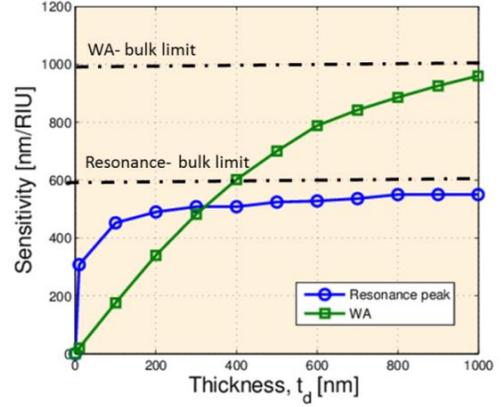

Figure 6. Numerically calculated sensitivity of the antenna (blue) and WA (green) resonances as a function of the analyte layer thickness for ambient medium with n=1.45.

In conclusion, we have presented and demonstrated a new sensing mechanism based on geometrical symmetry breaking of the WA resonances in nano-antenna slot arrays. In contrast to previously reported localized SPR sensors based on WA[30,31], our structure does not utilize plasmonic modes hybridization for obtaining a narrow resonance, but rather mode degeneracy lifting by geometrical symmetry breaking. In addition, since our structure comprises nano-vacancies in a metallic layer and not a metallic nano-particle, a substantial enhancement of the interaction between the localized electromagnetic field and the surrounding medium is obtained and contributes to the signal to noise of the device.

Sensitivity levels and FOMs exceeding 1000 nm/RIU and 200, respectively, were demonstrated experimentally, with very good agreement with the theoretical predictions at telecom wavelengths. Such a high FOM is mainly due to a narrow peak in the scattered spectrum emerge from the degeneracy removal of the leftwards and rightwards WA nulls under non-perpendicular illumination.

In contrast to the previously presented localized SPR sensors utilizing WA[30,31,41], the high FOM demonstrated here does not originate from the coherent coupling of the localized SPRs, but rather from residual response between two points of null scattering. Consequently, the main limitation on the attainable FOM in our structure is signal to noise ratio and not the intrinsic quality factor of the plasmonic modes, thus allowing for even higher FOMs. Moreover, additional improvement of the sensitivity and FOM can be attained through employment of advanced sensitivity enhancement techniques such as temperature differential methods [44] or even through a configuration change such as directing the light through the substrate, rather than through the analyte.

Finally, the concept can be easily converted to other desired wavelengths and RI ranges by modifying the slot dimensions and the array spacing, thus rendering it highly useful for numerous sensing applications including biosensing, water contaminant testing, and more. Furthermore, future research should be devoted to combining the sensing technique described above with the use of nano-transfer approaches[45] along with nano-imprinting lithography techniques[46] geared towards development of a low cost, high throughput fabrication technique.

## ASSOCIATED CONTENT

In the supporting information for this paper, we elaborate on the simulations and the fabrication methods employed for this work, as well as the experimental setup which was used for characterizing the angle-dependent spectral response of the Nano-slot antenna array (NSAA). In addition, we present the complete set of measurements done for various RI media and the corresponding sensitivity and Figure of merit (FOM), compare the results to the non-tilted NSAA, and describe further investigation of the Woods anomaly (WA) splitting phenomenon. Finally, we investigated the impact of the backplane reflector and the spacer thickness on the device sensitivity as well as the sensitivity as a function of thickness. This material is available free of charge via the Internet at http://pubs.acs.org

## AUTHOR INFORMATION

**Corresponding Author**

\* **kobys@eng.tau,ac.il**

**Author Contributions**

∥These authors contributed equally

Insert Table of Contents artwork here